# Conformable Nanowire-In-Nanofiber Hybrids for Low-Threshold Optical Gain in the Ultraviolet

*Alberto Portone[†,‡,§], Rocio Borrego-Varillas[#], Lucia Ganzer[#], Riccardo Di Corato[⊥], Antonio Qualtieri[∥], Luana Persano[†,‡], Andrea Camposeo[†,‡], Giulio Cerullo[#,*] and Dario Pisignano[†,¶,*]*

[†]NEST, Istituto Nanoscienze-CNR, Piazza S. Silvestro 12, I-56127 Pisa, Italy

[‡]NEST, Scuola Normale Superiore, Piazza S. Silvestro 12, I-56127 Pisa, Italy

[§]Dipartimento di Matematica e Fisica "Ennio De Giorgi", Università del Salento, Via Arnesano I-73100, Lecce, Italy

[#]IFN-CNR, Dipartimento di Fisica, Politecnico di Milano, Piazza L. da Vinci 32, I-20133 Milano, Italy

[⊥]Institute for Microelectronics and Microsystems, CNR-IMM, Campus Ecotekne, Via Monteroni, I-73100 Lecce, Italy.

[∥]Center for Biomolecular Nanotechnologies, Istituto Italiano di Tecnologia, Via Barsanti, I-73010 Arnesano (LE), Italy.

[¶]Dipartimento di Fisica, Università di Pisa, Largo B. Pontecorvo 3, I-56127 Pisa, Italy.

*Corresponding authors, giulio.cerullo@polimi.it, dario.pisignano@unipi.it

**Abstract**. The miniaturization of diagnostic devices that exploit optical detection schemes requires the design of light-sources combining small size, high performance for effective excitation of chromophores, and mechanical flexibility for easy coupling to components with complex and non-planar shapes. Here, ZnO nanowire-in-fiber hybrids with internal architectural order are introduced, exhibiting a combination of polarized stimulated emission, low propagation losses of light modes, and structural flexibility. Ultrafast transient absorption experiments on the electrospun material show optical gain which gives rise to amplified spontaneous emission, with threshold lower than the value found in films. These systems are highly flexible and can conveniently conform to curved surfaces, which makes them appealing active elements for various device platforms, such as bendable lasers, optical networks and sensors, as well as for application in bioimaging, photo-crosslinking, and optogenetics.

**Keywords**: organic-inorganic materials, nanocomposites, zinc oxide, electrospinning, amplified spontaneous emission, ultrafast transient absorption

The development of miniaturized and effective light sources with emission in the near ultraviolet (UV) is highly important for all those fields, including microanalysis through fluorescence spectroscopy, chemical sensing, and healthcare diagnostics, where molecular photo-excitation is involved.[1-4] Together with mechanical flexibility to have them conformed to different surfaces or lodged within different lab-on-chip platforms,[2,3,5,6] UV-active materials are desired to feature stimulated emission and optical gain, which is the basic prerequisite to use them in compact laser systems and photonic networks. Examples of UV-emitters include various inorganics (GaN, ZnS, ZnO, *etc.*) and their nanostructures grown through chemical vapour transport processes, colloidal synthesis, and other deposition methods.[7-12] Inorganic nanostructures, and especially nanowires (NWs), have been largely exploited within semiconductor-embedding optical cavities, plasmonic nanolasers, and heterojunctions.[7,9,12,13] However, the so-obtained devices are largely limited to planar systems or substrates for nanostructure growth, they are mechanically stiff, and unconformable to curved surfaces. Polymeric light-emitting materials and films,[3,14-17] instead, can be easily coupled with bendable substrates and are more versatile in terms of configurational motifs, although their optical performance and stimulated emission properties are generally much less appealing and stable in time compared to inorganics.

Hybrid approaches, inserting highly-efficient and stable oxide nanosystems in flexible polymer nanostructures and enabling high throughput processing, would combine advantages of UV light-emitting inorganics and plastic matrices, and potentially lead to materials with enhanced properties. For instance, in these systems light from UV nanoscale emitters could be efficiently coupled to polymer waveguides and transported along macroscale distances,[3,18] thus overcoming the high propagation losses typical of individual inorganic nanostructures[19] and enabling a much more versatile device interfacing. A few attempts at fabricating luminescent poly(ethylene terephthalate) or polyvinyl alcohol nanofibers embedding ZnO or Ge nanocrystals have been reported,[20,21] nonetheless the stimulated emission and optical gain properties of hybrid functional materials in the UV spectral range are still unexplored. In this work, we introduce hybrids encompassing UV-emitting ZnO NWs in polymer fibers and report on their optical gain properties. ZnO thin films and nanostructures are highly interesting due to their large exciton binding energy (~60 meV), efficient radiative recombination at room temperature, and large band gap (~3.3 eV), which provides very stable and narrow luminescence between 370 and 390 nm[7,22-24] and even potential for reliable polariton lasing at room temperature following embedment in microcavities.[25] In fact, previous methods based on *in situ* synthesis of ZnO nanocrystals from zinc acetate dehydrate precursors[20] have led to isotropic nanoparticles or grain aggregates in fibers, which are hard to engineer in order to obtain NWs in hybrids without increased surface roughness, that are instead desired to promote light transport and amplification. Here, the incorporation of ZnO NWs in transparent[3] poly(methyl methacrylate) (PMMA) electrospun filaments leads to a NW-in-fiber hybrid material with internal architectural order, combining optical gain, polarized stimulated emission, and waveguiding properties in an efficient system for flexible UV lasers operating at room temperature. The internal alignment of NWs promoted by electrospinning is found to be critically relevant in making ZnO NW-in-fiber hybrids capable to outperform counterparts doped by non-elongated ZnO nanocrystals, in terms of reduced optical losses as well as low amplified spontaneous emission (ASE) threshold.

The relevance of here reported results is twofold, *i.e.* at both fundamental and practical level. Ultrafast transient absorption measurements with high temporal resolution (20 fs) are presented for the ZnO NW-in-fiber hybrids. These measurements provide insight into the ultrafast non-equilibrium physics of ZnO NWs in the hybrid material, enabling the observation of carrier scattering with optical phonons and of a shift in absorption due to photogenerated charges (Burstein-Moss effect), thus providing insight into the ultrafast non-equilibrium physics of ZnO NWs in the hybrid material. In addition, these hybrids can lead to the definition of design rules for miniaturized lasers, bio- and chemical sensors, and optical networks embedding UV-emitting materials and nano-components. Indeed, the optical gain in the hybrid system incorporating elongated nanoparticles can benefit from having them aligned in electrospun polymers,[26] promoting enhanced forward scattering[27,28] and channeling photons along the longitudinal axis of the fibers. The latest applications benefiting from the advanced combination of properties of NW-in-fibers include wearable components for healthcare monitoring and biometric recognition, and devices for optogenetics, photo-chemical cross-linking of proteins, localized Calcium imaging, and immunological modulation.[29-32]

### Results and Discussion

The used ZnO nanomaterials, including non-elongated nanoparticles (NPs) and NWs, are initially inspected by transmission electron microscopy (TEM, Figure 1). NPs generally exhibit diameters below 100 nm, whereas NWs have transverse section of about (40±15) nm and various lengths with maximum of the order of 1 μm, leading to an aspect ratio (length/diameter) below 25. For NWs with this transversal size and length, one could expect a minor influence of surface effects on optical properties.[33,34] In addition, for this range of sizes, quantum confinement effects do not play a significant role and the electronic and optical properties of the nanomaterials resemble those of the bulk.[35,36] The dispersion in size of ZnO NPs and NWs is not found to affect significantly the fabrication process of doped fibers. In fact, in order to embed these systems in polymer filaments with limited aggregation and precipitation, dispersions of NPs and NWs are prepared with a mixture of chloroform and ethanol (1:1 v:v) following extensive solubility tests (see Figure S1 in the Supporting Information file). Dimethylformamide (DMF) is also added to reduce the overall solvent evaporation rate and avoid needle clogging during electrospinning.

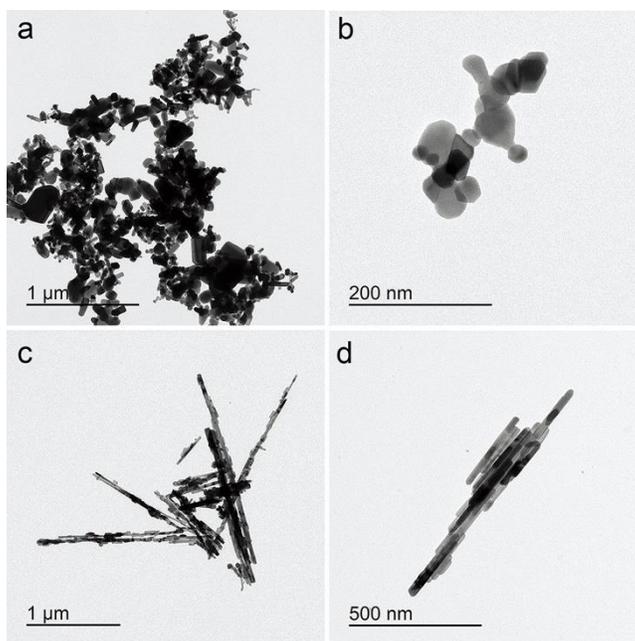

**Figure 1.** TEM images of used ZnO NPs (a,b) and ZnO NWs (c,d) at different magnifications. Both NPs and NWs easily form clusters before imaging.

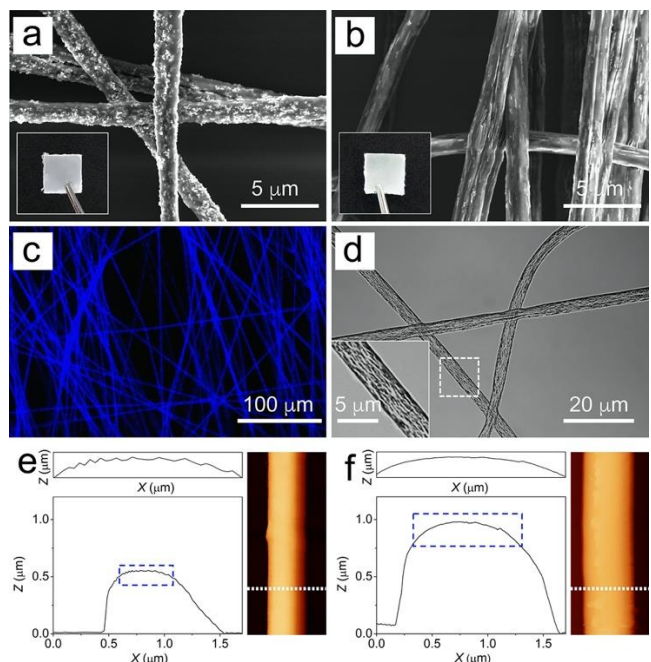

**Figure 2.** (a,b) SEM micrographs of fibers doped with ZnO NPs (a) and ZnO NWs (b). Insets show the photographs of the fibrous samples. (c) Fluorescence image of electrospun fibers doped with ZnO NPs. (d) ZnO NWs-doped fibers imaged by transmitted light microscopy Inset: magnification of the sample area highlighted by a dashed square box. (e,f) Profiles of exemplary fibers incorporating ZnO NPs (e) and ZnO NWs (f), along the dashed line in the corresponding AFM images shown on the right of the graphs. Top panels: magnifications of the profile regions in the blue-dashed rectangles (top regions of fibers). Relative weight ratio (NPs/polymer or NWs/polymer) in the electrospinning solution: $\chi$ = 30%.

Exemplary scanning electron microscopy (SEM) images of ZnO NPs- and NWs-doped fibers, spun from solutions with nanomaterial/polymer ratio ($\chi$) as high as 30% in weight, are shown in Figure 2a,b. The corresponding diameter distributions, for fibers with different doping, are displayed in Figure S2. The average diameter of the fibers is found to slightly increase upon increasing $\chi$, up to a maximum of about 1.8 μm, which is associated with the varied rheology of the electrospun solution.[37] Fibers are fluorescent, and the uniform emission signal collected from different segments of each filament (Figure 2c) as well as TEM imaging (Figure S3) indicate that dopants are evenly distributed along the length of the filaments. Overall, fiber diameters are well above the cut-off value[38] ($d_{cut-off}$) for waveguiding at their emission wavelength ($d_{cut-off} \cong$ about 160 nm, as estimated in the Supporting Information, Figures S4 and S5), which allows the intensity of the guided light modes to be effectively confined.

In addition, the SEM analysis strongly suggests a largely uniaxial alignment of NWs along the fiber length, which is also supported by optical transmission micrographs (Figure 2d). This alignment also involves NWs close to the fiber surfaces, thus significantly affecting external roughness of NW-in-fiber samples. Indeed, the surface of NW-in-fibers is significantly smoother (with root mean squared roughness, $R_{q\_NWs}$ = 19.7 nm) compared to that of NPs-doped fibers ($R_{q\_NPs}$ = 43.2 nm) where clustered particles may partially protrude out from the polymer surface, as measured by atomic force microscopy (AFM, Figure 2e,f). The alignment of ZnO NWs along the fiber longitudinal axis is also supported by TEM measurements on the hybrid filaments (Figure 3a,b), as well as by SEM performed by utilizing backscattered electrons (Figure 3c). The maximum misalignment angle is found to be as low as 10° (Figure 3d). Such tight control of the NWs orientation is obtained during fiber generation, and alignment promoted by the dynamics of elongated particles within the initially sinklike flow at the onset of the electrospinning process as well as within the stretching and thinning electrified solution.[26,39] Internal ordering of NWs within each polymer filament impacts on ASE properties, enhancing them as better detailed in the following. Indeed, the alignment of NWs is expected to minimize the propagation losses for radiation guided along the longitudinal axis of the hybrid fibers, because of the dominant forward component[27,28] of light-scattering at the ZnO/polymer interface (Figure S6) thus promoting waveguiding.

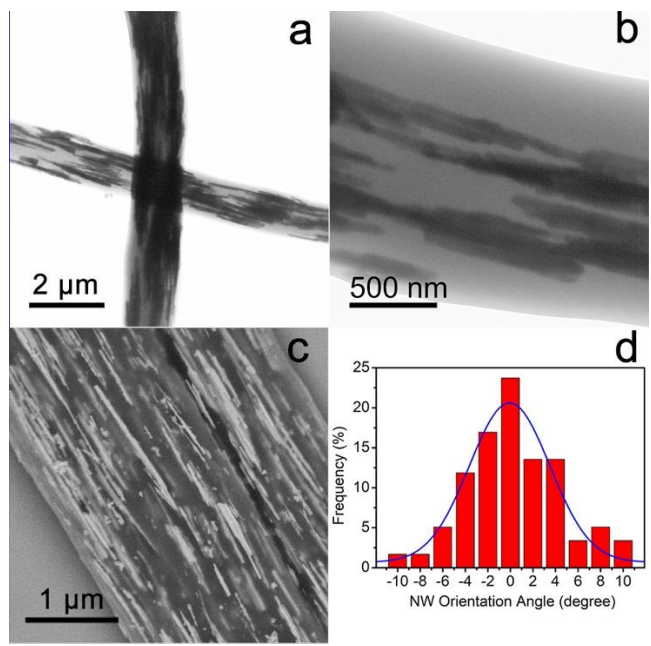

**Figure 3.** TEM (a,b) and SEM (c) micrographs of composite fibers with ZnO NWs. $\chi = 30\%$. (d) Distribution of the orientation angles of the ZnO NWs with respect to the fiber longitudinal axis. 0° corresponds to the direction parallel to fiber length.

Reference absorption measurements performed on the hybrid materials highlight the typical features of ZnO,[40] with a single narrow peak at 374 nm due to the excitonic transition superimposed to a light-scattering background (Figure 4a). Photoluminescence (PL) properties are also analyzed on both spin-cast films and mats of fibers doped with either NPs or NWs, as displayed in Figure 4b and 4c. A bright UV emission band is measured, as typical of exciton recombination in ZnO, which peaks at 383 nm and 384 nm for NPs and NWs, respectively. Here, the spontaneous emission of the hybrid fibers (red lines in Figure 4b,c) is largely analogous to that of the corresponding films (black lines), with only minor red-shifts (1-3 nm) of the peak wavelengths. This effect can be attributed to the slightly enhanced self-absorption, that is likely to be more effective in fibers due to their waveguiding properties, since these favor interactions of photoemitted light with embedded active NPs/NWs over longer effective distances. The insets of Figure 4b,c show a magnification of the emission spectra in the region between 450 nm and 700 nm. Here, a weak fluorescence is measured (one-two orders of magnitude below the intensity of the main emission band in the UV), generally associated to ZnO surface states and defects such as oxygen and zinc vacancies and interstitials.[41-43] Furthermore, the comparison of the PL and ASE intensity from fibers electrospun from solutions with different NW/polymer ratios, measured under identical excitation and collection configurations, shows a brighter UV emission for $\chi = 30\%$ (Figure S7). A lower PL intensity is measured for fibers obtained from $\chi = 50\%$ solutions, which is likely related to higher self-absorption of the emitted light as also evidenced by the corresponding red-shift of the emission peak (Figure S7a). ASE is found only in fibers from $\chi = 30\%$ solutions, with other hybrids suffering from either low content of the active medium ($\chi = 10\%$) or higher self-absorption ($\chi = 50\%$, Figure S7b).

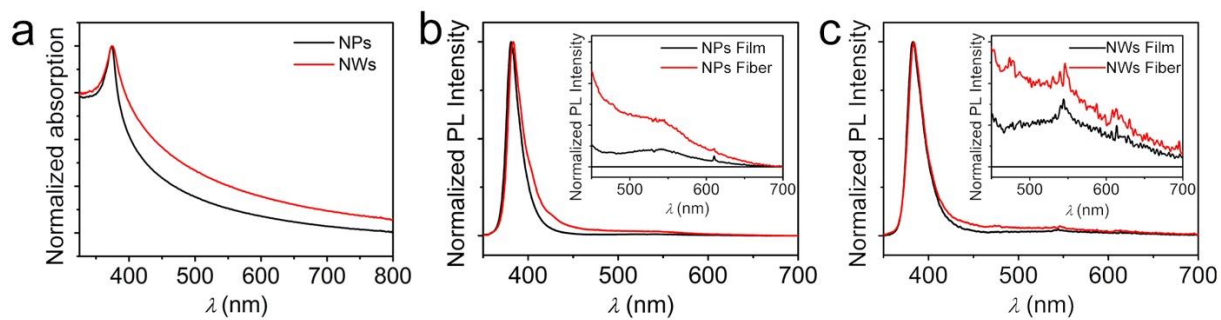

**Figure 4.** (a) Normalized absorption spectra of PMMA films doped with ZnO NPs (black line) and NWs (red line). (b,c) Normalized PL emission of films (black lines) and fibers (red lines) doped with ZnO NPs (b) and NWs (c). Insets: magnified PL spectra in the visible spectral range.

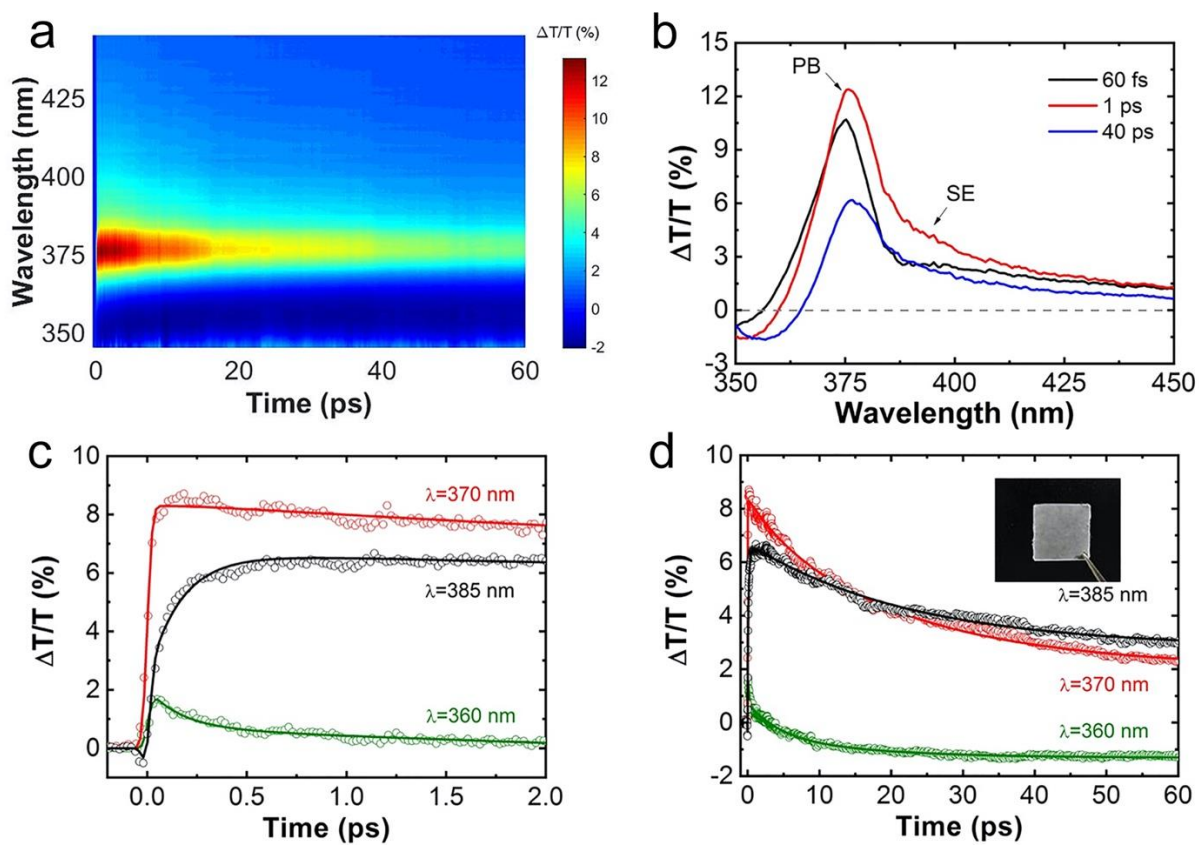

**Figure 5.** Femtosecond pump-probe spectroscopy of fibers doped with ZnO NWs. (a) 2D $\Delta T/T$ map as a function of probe wavelength and pump-probe delay; (b) $\Delta T/T$ spectra at selected pump-probe delays. (c,d) $\Delta T/T$ dynamics at selected probe wavelengths for short (c) and long (d) delays. Pump-probe measurements are performed on samples from solutions with $\chi$ =10% -see picture in the inset of panel (d)-, to avoid over-attenuation of transmission and spectral artifacts due to light-scattering from the bulk nanocomposite.

In order to study the non-equilibrium optical response of the UV hybrid materials, we perform femtosecond pump-probe spectroscopy with excitation at 335 nm (3.7 eV) and broadband detection spanning from 330 to 550 nm. A 20 fs pump pulse is used to photoexcite carriers well above the bandgap of ZnO, whereas a delayed white-light-continuum probe pulse measures the differential transmission ($\Delta T/T$) spectrum as a function of time delay (see Methods). Figure 5a shows the $\Delta T/T$ map, as a function of probe wavelength and pump-probe delay, for hybrid fibers embedding NWs. Figure 5b shows $\Delta T/T$ spectra at selected pump-probe delays, whereas panels 5c and 5d show $\Delta T/T$ dynamics at selected probe wavelengths. The corresponding results for fibers embedding NPs, shown in Figure S8, display a similar behavior. The excitation fluence is 250 μJ/cm$^2$, which corresponds to a carrier density of $6.78 \times 10^{16}$ cm$^{-3}$ (see Supporting Information), more than one order of magnitude below the threshold for the Mott transition in ZnO.[44] The $\Delta T/T$ spectrum at 1 ps pump-probe delay (Figure 5b) shows a positive signal at 365-380 nm, which is assigned to photobleaching (PB) of the excitonic transition (*i.e.* reduced absorption of the sample following photoexcitation), as expected on the basis of the absorption spectra reported in Figure 4a. Importantly, a positive $\Delta T/T$ signal is also observed for wavelengths in the 380-400 nm range, that corresponds to the UV spectral region of the excitonic PL (Figure 4b,c). Such positive signal can be attributed to stimulated emission (SE), suggesting the presence of net optical gain[45] in the material. Finally, a positive signal is also observed at longer wavelengths, up to 450 nm, and is attributed to a combination of PB and SE arising from ZnO surface states and defects. The $\Delta T/T$ data are subjected to global analysis[46] and the corresponding decay associated spectra (DAS) are shown in Figure S9. DAS1 has a time constant ∼ 240 fs, and it can be attributed to the relaxation of the photogenerated hot carriers through coupling with optical phonons.[47] DAS1 shows the build-up of the SE band peaking at 380 nm (as confirmed by the 385 nm dynamics in Figure 5c) as well as the build-up of a photoinduced absorption (PA) band peaking at 360 nm. This PA band[48,49] can be tentatively assigned to the Burstein-Moss effect. DAS2, with a weak amplitude and a 5.6 ps time constant, may be attributed to further carrier thermalization *via* equilibration of optical and acoustic phonons. DAS3 corresponds to a decay of the PB and SE signals and has a time constant of 38.7 ps, in good agreement with previous time-resolved photoluminescence studies of ZnO,[50,51] which attributed it to the decay of free excitons. Finally, the long-lived DAS4 corresponds to the decay of excitons bound to defects and band-edge states. Overall, the ultrafast studies clearly suggest that the hybrid fibers with both NPs and NWs can be promising flexible gain media for UV lasing.

To demonstrate the use of these systems in mirrorless laser sources, ASE measurements are carried out by optically-pumping uniaxially aligned arrays of fibers with long-pulse (ns) excitation at $\lambda = 355$ nm. The pump light is polarized parallel to the fiber length and shaped as a stripe with size $10 \times 0.2$ mm$^2$, which is aligned parallel to fibers. ASE is generated by the amplification of spontaneously emitted photons by stimulated emission under intense optical excitation,[52] a process that can be also assisted by the waveguiding of light along the polymer fibers with ZnO NWs. ASE displays a characteristic spectral line-narrowing upon increasing the excitation intensity. Such line-narrowing occurs in the spectral region of maximum net gain of the active material, that is related to the stimulated emission cross-section and to the spontaneous emission rate of the active material, as well as to the propagation losses and the properties of light modes in the waveguide. No ASE is found in hybrid filaments doped with ZnO NPs, evidencing the critically important role of waveguiding in assisting ASE along polymer matrices and nanocomposites. Indeed, upon NP-doping the rough fiber surface (Figure 2e) disfavors efficient waveguiding and consequently amplification of light emitted by ZnO. Rayleigh scattering from surface defects is likely to be a significant source of optical losses, whose magnitude coefficient per unit propagation length ($\alpha$) can be estimated by using a formulation of the Rayleigh criterion originally derived for slab waveguides:[53]

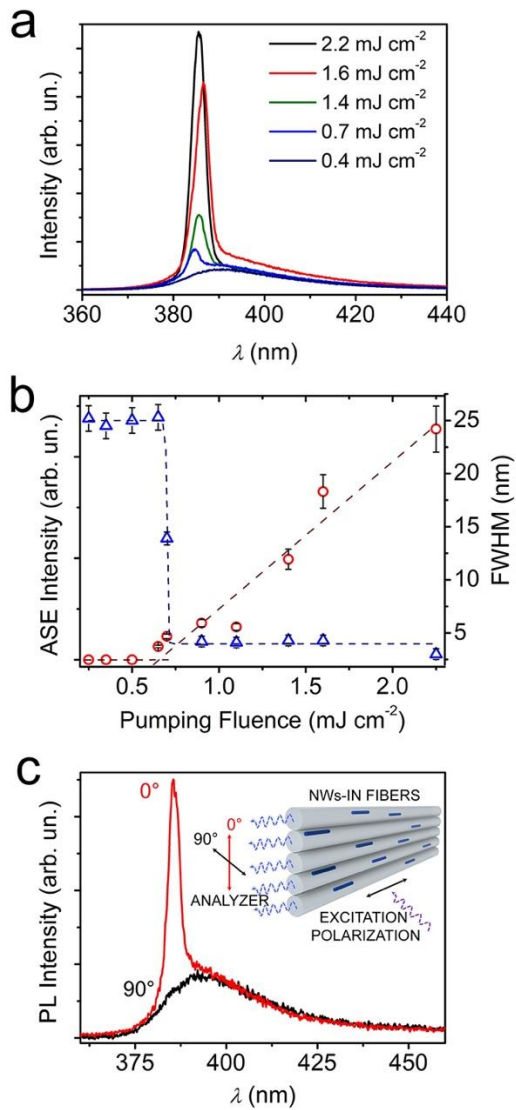

**Figure 6.** (a) PL spectra of NW-in-fibers for different pumping fluence, highlighting the ASE peak. (b) ASE intensity (red circles, left vertical scale) and FWHM of the spectra (blue triangles, right vertical scale) as a function of the pump fluence. (c) Polarization dependent ASE intensity from NW-in-fibers. The emitted light is collected from the cut end of the fibers by an analyzer with polarization parallel (0°, red line) and perpendicular (90°, black line) to the sample surface (scheme in the inset). $\chi$ = 30%.

$$\alpha = (4\pi/\lambda_{ASE})^2 \times (\cos^3\theta_i /2\sin\theta_i) \times [R_q{}^2/(D+L)]. \tag{1}$$

In Eq. (1), $\theta_i$ indicates the angle of incidence, on the fiber surface, of the optical mode propagating along the waveguide with wavelength $\lambda_{ASE}$ (for PMMA waveguides, one has $\theta_i$ >75°), $D$ is the fiber diameter, $L$ is the mode penetration depth into the environment surrounding the fiber ($L << D$ for $D$ of the order of micrometers). This leads to an optical loss coefficient associated to surface Rayleigh scattering from ZnO NP-doped fibers about five times larger than the corresponding value for NW-in-fibers, which may explain the lack of ASE in the system embedding non-elongated ZnO NPs.

Figure 6 shows the ASE spectra that are instead obtained by ZnO NW-in-fibers, at various pump fluences, and the corresponding light-light (*L-L*) plot. The typical spontaneous emission of ZnO NWs, with a spectrum peaking at about 384 nm, can be observed at low excitation fluence (≤500 μJ cm$^{-2}$) in Figure 6a. Upon increasing the excitation fluence above about 600 μJ cm$^{-2}$, a sharp peak at about 382 nm is initially observed, which very rapidly increases in intensity as typical of ASE (circles in Figure 6b). In fact, due to the multiple mechanisms at play, the peak wavelength of ASE does not coincide with the peak wavelength of the spontaneous emission. In particular, the ASE peak is found to correspond to the spectral region of excitonic recombination, while ZnO surface states and defects contributing to the broader spontaneous emission do not appear to undergo stimulated emission assisted by waveguiding in the electrospun polymer fibers. We also find a blueshift (9 meV) of the ASE peak energy for very high excitation densities (from 1.6 up to 2.2 mJ cm$^{-2}$), which is a signature of the increasing screening of the exciton binding energy, as a consequence of the increased density of electron-hole pairs created upon photoexcitation.[54,55] Such effect competes with the band gap renormalization that typically red-shifts the gain profile as the excitation intensity is increased,[54] as here found for excitation fluences up to 1.6 mJ cm$^{-2}$. Correspondingly, the full width at half maximum (FWHM) of the PL spectrum decreases from 25 nm to about 4 nm (triangles in Figure 6b) upon increasing the excitation fluence. Although ASE is an inherently thresholdless phenomenon,[56] an experimental effective threshold is usually introduced to quantitatively describe gain narrowing, defined as the halfway fluence between the spontaneous emission and the narrowed ASE linewidth regimes. Here we measure a threshold of 650 μJ cm$^{-2}$ for ZnO NW-in-fibers, that is significantly lower than values reported for other hybrid systems based on ZnO nanomaterials such as polydimethylsiloxane nanocomposite films and zinc oxide-silica systems (generally in the range 2-4 mJ cm$^{-2}$).[57,58] Various effects might contribute in enhancing ASE features in UV light-emitting fibers with respect to films, such as the alignment of ZnO NWs along the fiber length which leads to more effective photo-excitation, as well as waveguiding along the longitudinal axis of the aligned filaments which directly promotes ASE.[59]

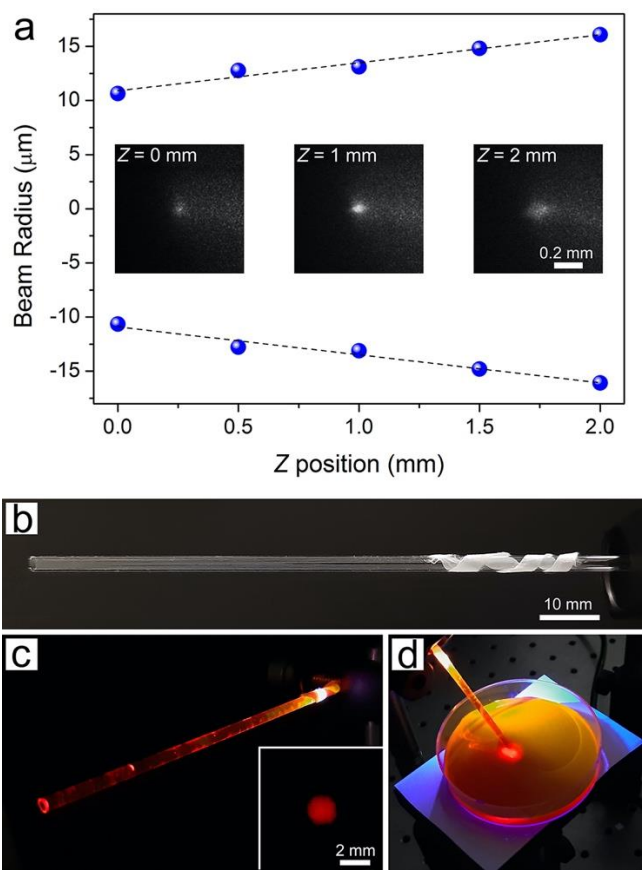

**Figure 7.** (a) Measured ASE beam divergence. ASE beam radius (dots) *vs*. detector distance, *Z*. Inset: micrographs of the emission spot measured at different distances. (b-d) Examples of hybrid NW-in-fibers coupled with non-planar surfaces and optical systems. (b) NW-in-fiber mat wrapped around the curved surface of a capillary glass tube. (c,d) The UV light from NW-in-fibers excites a viscous Rhodamine 6G solution in the tube. The light emitted by the dye is then guided along glass and the viscous fluid, and generates a ~2 mm spot (inset of panel c) at the capillary end for liquid sample illumination (d).

In addition, ASE from NW-in-fibers is found to be polarized along the direction parallel to the sample surface ('0°' curve in Figure 6c). This can be explained by the higher propagation losses experienced by light with polarization component perpendicular to the sample surface ('90°' curve in Figure 6c), due to more intense outcoupling into the quartz substrate underneath.[60,61] The linearly polarized ASE could be exploited for significantly improving the signal-to-noise ratio in diagnostic chips working through fluorescence excitation.[62] To this aim, also noteworthy is the high directionality of the emitted light. We analyze the ASE beam profile of aligned NW-in-fibers at various distances from the tip (Figure 7a), evidencing a divergence of 5 mrad, which is lower than that reported for other polymer light-emitting systems in the form of film[63] or fibers[3] (Table S1). Data for ZnO NW-doped films are displayed in Figure S10. They exhibit a worse ASE performance, with a more than one order of magnitude higher threshold, which we attribute to the disordered internal configuration of the dopants.

Finally, the NW-in-fiber hybrids are conformable enough to be wrapped around curved surfaces with mm-scale curvature radius such as capillary glass tubes (Figure 7b), thus adapting to specific excitation schemes (Figure 7c,d). For instance, the UV light emitted by the optically-pumped hybrids here in contact with the glass tube is able to excite chromophores (Rhodamine 6G dye) flowing in the capillary (Figure 7c). The light emitted by the dye can then be guided along the same capillary for centimeters, until it generates a well-defined exit spot (inset of Figure 7c) for remote sample illumination at the chromophore wavelength, as shown by a solution in the plate photographed in Figure 7e.

A summary of the emissive properties for ZnO NW-in-fibers is reported in Table S1 in the Supporting Information, comparing achieved performance with those of other miniaturized, optically excited UV light-emitting materials. Overall, the occurrence of ASE under long-pulse pump conditions, combined with the photo- and chemical stability of embedded ZnO nanostructures, the high directionality of the emitted light and the achieved flexibility of the hybrid material, make NW-in-fibers excellent and versatile active media for flexible UV lasers that can be coupled with curved optical components or integrated in miniaturized diagnostics and microfluidic devices.

**Conclusion**

In summary, we have introduced light-emitting NW-in-fiber hybrid materials with optical gain in the UV and highly favourable combination of properties. We performed an extensive optical characterization of the electrospun fibers. High time-resolution pump-probe spectroscopy in the UV detected stimulated emission in the 380-400 nm wavelength range, while line narrowing in this spectral range provided a signature of ASE, with onset for excitation fluence of 650 μJ cm$^{-2}$. ASE in the fibers with internal order of ZnO NWs, leading to more effective optical behaviour, features a lower excitation threshold compared to other nano-hybrid materials and components based on ZnO nanocrystals. These properties, obtained in flexible and conformable fiber mats, make these materials highly interesting for various applications in sensing and light-emitting devices, bioimaging and photo-crosslinking, as well as in network and bendable microlasers. Upon engineering the strength of the ZnO/polymer interface through proper surface modifiers or cross-linking of the organic matrix, one could also aim to obtain effective strain and load transfer to the inorganic component,[64,65] a perspective highly interesting for opto-mechanical coupling[66] and to exploit the piezoelectric properties of NWs in the hybrid material.

**Methods**

*Hybrid NW-in-fibers.* ZnO NWs (manufacture code 773999) and ZnO NPs (manufacture code 544906) with high purity[67-69] are purchased from Sigma Aldrich. X-ray diffraction (XRD) data on these materials[69-72] indicate typical hexagonal wurtzite structures with high crystallinity. PMMA (average molecular weight~120,000 Da), and chloroform are purchased from Sigma Aldrich. Ethanol is purchased from Fluka and DMF is purchased from Carlo Erba. Solutions for electrospinning are prepared by 250 mg mL$^{-1}$ of PMMA dissolved in a mix of chloroform, ethanol and DMF with volume ratio 4:4:1 and ZnO NPs or NWs added at a relative weight ratio ($\chi$) between 10% and 50% with respect to the polymer. The solution is loaded in a syringe with a 27 gauge needle and injected through the needle at constant flow rate of 1 mL h$^{-1}$ by a microprocessor dual-drive syringe pump (33 Dual Syringe Pump, Harvard Apparatus Inc., Holliston, MA). A voltage of 12 kV (XRM30P, Gamma High Voltage Research Inc., Ormond Beach, FL) is applied to the spinneret, and fibers are collected at distance of 10 cm from the needle on a metallic collector biased at −6 kV. Either a plate or a rotating collector (disk with 0.8 cm width, 8 cm diameter, 4000 rpm) are used to obtain randomly-oriented or aligned fibers, respectively.

*Morphological and optical characterization.* The morphology of the hybrid fibers is investigated by SEM, TEM and AFM. For SEM analysis, fibers are collected on Si substrates and images acquired by a FEI Helios 600i system, operating at acceleration voltages of 3-5 kV. TEM images are acquired with a JEOL JEM 1011 microscope operating at accelerating voltage of 100 kV. For particle inspection, few drops of dispersions are drop-casted onto 300 mesh carbon-coated grids, and the residual solvent is allowed to evaporate at room temperature. For fibrous samples, 400 mesh copper grids are temporarily fixed on the electrospinning collector by carbon tape, and few fibers deposited onto them in a short time lapse ($\cong$ 10 s) for subsequent imaging. All images are acquired and processed by Gatan Digital Micrograph 2.3 native software. AFM measurements are carried out by a Multimode system equipped with a Nanoscope IIIa electronic controller (Veeco Instruments). The fiber topography is analyzed in tapping mode, utilizing Si cantilevers with 190 kHz resonance frequency and with tips featuring a nominal radius of 8 nm, length in the range 10-15 μm and apex angle of 22.5°. The width and height of the cross-sectional profile of the NW-in-fiber (Figure 2f), calculated upon deconvolution with the profile of the used AFM tip, are indicative of an almost circular cross-section (~1 and ~0.9 μm, respectively).

Fluorescence and transmitted light micrographs are collected by means of an inverted microscope Eclipse Ti (Nikon), using a 20× (numerical aperture, $N.A.$ = 0.50, Nikon) and a 60× ($N.A.$ = 1.40, Nikon) objective. Fluorescence images are acquired by exciting samples with a Hg lamp and collecting the emitted light with a digital camera DS-Ri1 (Nikon). Maps of the light transmitted by samples are collected by scanning a focussed Argon ion laser ($\lambda$ = 488 nm). Absorption spectra are collected by a double beam ultraviolet-visible spectrophotometer (Perkin Elmer).

*PL, ASE, and pump-probe experiments*. PL and ASE measurements are carried out by exciting NW-in-fibers, deposited on quartz substrates, by the 3$^{rd}$ harmonic of a ns Nd:YAG laser (Quanta-Ray INDI Spectra-Physics), with emission wavelength at $\lambda$ = 355 nm and a repetition rate of 10 Hz. The propagation direction of the incident pumping beam is set perpendicular to the sample surface, and the emission is collected from the sample edge, namely at 90° with respect to the excitation beam, through a fiber-coupled imaging spectrometer (iHR320, Jobin Yvon) equipped with a charge coupled device (Simphony, Jobin Yvon). Beam divergence measurements are performed on aligned fibers excited above threshold. A Si CCD is used to measure the spatial profile of the emitted beam, and the divergence is estimated plotting the measured beam diameter *vs.* the detector position, and evaluating the half angle corresponding to the asymptotic variation of the beam radius along the propagation direction.

Femtosecond pump-probe experiments are performed starting from an amplified Ti:sapphire system (Libra, Coherent) generating 4-mJ, 100-fs pulses at 800 nm central wavelength and 1 kHz repetition rate. A fraction of the laser output is used to pump a non-collinear optical parametric amplifier (NOPA), which generates sub-20-fs visible pulses at 540 nm. Sum-frequency generation of the NOPA output with the 800-nm pulse in a 50-μm-thick type II β-barium borate crystal produces the sub-20-fs pump pulses with spectrum centred at 335 nm.[73] Another fraction of the laser output is focused in a 3-mm-thick $CaF_2$ plate, to generate a white-light continuum covering the 330-650 nm wavelength range, used as broadband probe pulse. Pump and probe pulses, both polarized parallel to the direction of fibers, are non-collinearly focused on the sample and the transmitted probe is sent to a spectrometer capable of single-shot detection at the 1 kHz laser repetition rate. For each measurement, the transmittance of the probe pulse is measured with ($T_{pump\_ON}$) and without ($T_{pump\_OFF}$) a preceding excitation pulse. The relative changes in transmittance are expressed as $\Delta T/T$ (%) = [($T_{pump\_ON}$ - $T_{pump\_OFF}$)/ $T_{pump\_OFF}$]×100. A detailed description of the femtosecond pump-probe system can be found in Ref. 74. Global analysis of the pump-probe data is performed using the Glotaran software obtaining the DAS and the corresponding time constants.[75]

ASSOCIATED CONTENT

**Supporting Information**

Additional details and calculations on ZnO NPs solubility, diameter distribution of hybrid fibers, TEM analysis and estimation of the ZnO/PMMA volumetric ratio, calculated waveguiding and scattering properties, PL and ASE emission, estimation method for the photoexcited carrier density, femtosecond pump-probe spectroscopy of fibers with ZnO NPs, decay associated spectra, and ASE spectra of NW-doped films. The Supporting Information file (PDF) is available free of charge at https://pubs.acs.org/journal/ancac3.

AUTHOR INFORMATION

**Corresponding Authors**

Dario Pisignano, email: dario.pisignano@unipi.it

Giulio Cerullo, email: giulio.cerullo@polimi.it

**Author Contributions**

The manuscript was written through contributions of all authors. All authors have given approval to the final version of the manuscript.

ACKNOWLEDGMENT

The research leading to these results has received funding from the European Research Council under the European Union's Seventh Framework Programme (FP/2007-2013)/ERC Grant Agreement n. 306357 (ERC Starting Grant "NANO-JETS") and from the Italian Minister of University and Research PRIN 2017PHRM8X and PRIN 201795SBA3 projects. D.P. also acknowledges the support from the project PRA_2018_34 ("ANISE") from the University of Pisa. A. P. and A. C. acknowledge funding from the European Research Council under the European Union's Horizon 2020 Research and Innovation Programme (Grant Agreement no. 682157, ''*x*PRINT'').

# Supporting Information

# Conformable Nanowire-In-Nanofiber Hybrids for Low-Threshold Optical Gain in the Ultraviolet

*Alberto Portone[†,‡,§], Rocio Borrego-Varillas[#], Lucia Ganzer[#], Riccardo Di Corato[⊥], Antonio Qualtieri[∥], Luana Persano[†,‡], Andrea Camposeo[†,‡], Giulio Cerullo[#,\*] and Dario Pisignano[†,¶,\*]*

[†]NEST, Istituto Nanoscienze-CNR, Piazza S. Silvestro 12, I-56127 Pisa, Italy

[‡]NEST, Scuola Normale Superiore, Piazza S. Silvestro 12, I-56127 Pisa, Italy

[§]Dipartimento di Matematica e Fisica "Ennio De Giorgi", Università del Salento, Via Arnesano I-73100, Lecce, Italy

[#]IFN-CNR, Dipartimento di Fisica, Politecnico di Milano, Piazza L. da Vinci 32, I-20133 Milano, Italy

[⊥]Institute for Microelectronics and Microsystems, CNR-IMM, Campus Ecotekne, Via Monteroni, I-73100 Lecce, Italy.

[∥]Center for Biomolecular Nanotechnologies, Istituto Italiano di Tecnologia, Via Barsanti, I-73010 Arnesano (LE), Italy.

[¶]Dipartimento di Fisica, Università di Pisa, Largo B. Pontecorvo 3, I-56127 Pisa, Italy.

*Corresponding authors, giulio.cerullo@polimi.it, dario.pisignano@unipi.it

# 1. Solubility Tests

To identify a good solvent, able to disperse the nanomaterials and to avoid aggregation and precipitation during the solution preparation and the electrospinning process, extensive solubility tests have been performed by mixing ZnO nanoparticles with common solvents and mixtures of them. To this aim, 50 mg mL$^{-1}$ of NPs have been dispersed in (1) acetone, (2) chloroform, (3) a mixture of chloroform and dimethylformamide (ratio 1:1 volume/volume - v/v), (4) ethanol, (5) methanol, (6) a mixture of chloroform and ethanol (ratio 1:1 v/v). The NPs have been dispersed by an ultrasonic bath for 1 hour, and the resulting dispersions stored for 24 hours. As shown in Figure S1, after one day the NP-suspension in ethanol, methanol and in the mix of ethanol and chloroform is stable. On the contrary, NPs precipitate in absence of alcohols, forming a sediment.

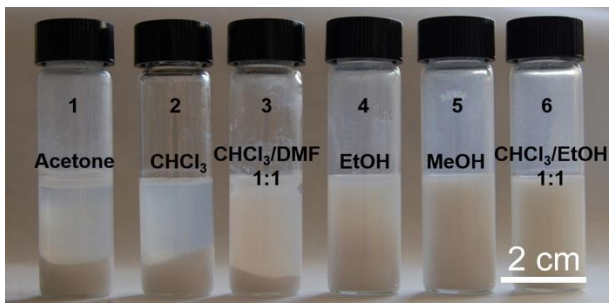

**Figure S1.** Photograph of ZnO NPs dispersed in various solvents: acetone (1), chloroform (2), chloroform and dimethylformamide with 1:1 (v/v) ratio (3), ethanol (4), methanol (5), chloroform and ethanol with 1:1 (v/v) ratio (6). Particle concentration: 50 mg mL$^{-1}$. Photograph acquired after 24 hs following dispersion by sonication. The suspensions in ethanol, methanol and in the mix of ethanol and chloroform are stable. On the contrary, NPs precipitate in absence of alcohols and form sediment. The chloroform/ethanol mixture is chosen for electrospinning experiments due to the very good solubility of PMMA in chloroform.

## 2. Size distribution

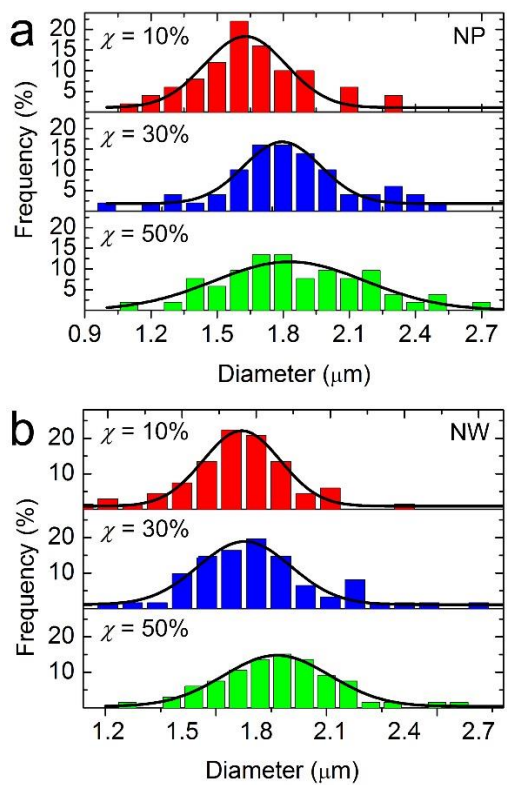

**Figure S2.** (a,b) Diameter distribution of fibers doped with ZnO NPs (a) and ZnO NWs (b), electrospun from solutions at various ZnO/polymer relative weight ratio ($\chi = 10\%$, 30%, and 50%). The superimposed black lines are Gaussian fits to experimental data.

## 3. TEM analysis

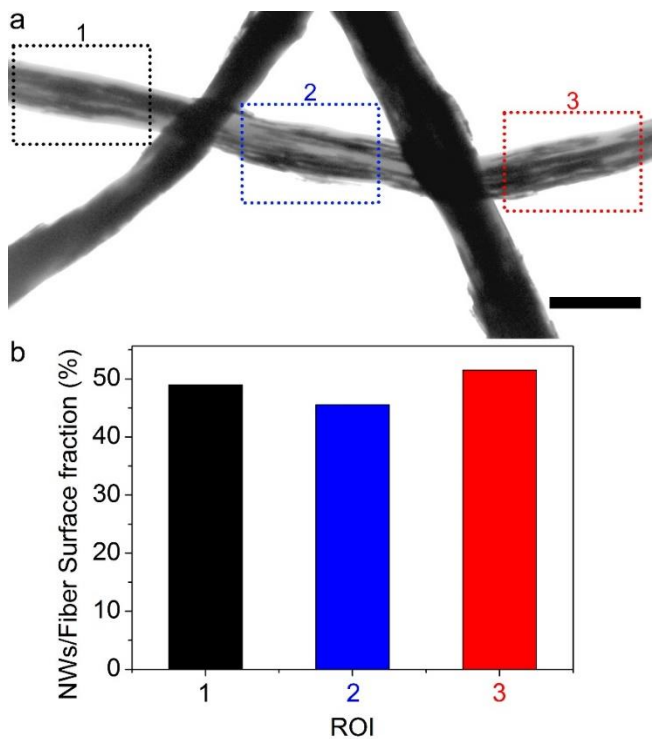

**Figure S3.** (a) TEM micrograph of fibers doped with ZnO NWs. $\chi = 30\%$. Scale bar: 2 μm. (b) Projected, shadowed area corresponding to NWs in the three regions of interest (ROIs) shown in (a).

TEM imaging was exploited to give an estimation of the ZnO/PMMA volumetric ratio in the hybrid fibers. An example of a TEM micrograph of a single hybrid fiber is shown in Figure S4. NWs are visible as darker areas with different signal intensities that can be attributed to two or more overlapping NWs along the path of the electron beam. Brightness and contrast of the image were set to highlight as much as possible the presence of NWs in the polymer matrix. The volume occupied by the NWs can be estimated by considering the area, $A_1$, of the darker regions as due to packed and uniaxially aligned NWs, with length 1 µm and diameter 40 nm. The ratio between $A_1$ and the cross-sectional area of a single NW ($A_{NW} \cong 2L_{NW} \times r_{NW}$, where $L_{NW}$ and $r_{NW}$ are the length and radius of the ZnO NW) gives the number of NWs in the area $A_1$ ($N_1 = A_1/A_{NW}$) and allows their total volume ($V_1 = N_1 \times V_{NW} = N_1 \times L_{NW} \times \pi r_{NW}^2$) to be calculated. The extension of $A_1$ is shown in Figure S4b, where a thresholding method was applied to the original image to remove the signal due to the polymer matrix. Briefly, in a grayscale image (whose scale, $s$, ranges from 0, that corresponds to black, to 255, that corresponds to white) the method set to white the intensity of those pixels whose $s$ value is higher than a fixed value (threshold value) and to black the intensity of those pixels whose $s$ is lower than the threshold value. The black area of Figure S4b was measured and considered as $A_1$. Such estimation can be improved by considering that in some areas there is an overlap of two NWs, as highlighted by the red arrow in Figure S4a. To account for this effect, the TEM micrograph was processed by selecting a different threshold value, *i.e.* a value lower than the one used for estimating $A_1$ and also lower than the $s$ value for a single NW, as the one highlighted by the blue arrow in Figure S4. Indeed, the area highlighted by the blue arrow is black in Figure S4b, and it is included in the calculation of the area of $A_1$, while it is white after the second thresholding (Figure S4c). In this way, the extension of an area $A_2$ (black area in Figure S4c) can be utilized to estimate the number of NWs, $N_2$, whose shadowed regions overlap with those included in the area $A_1$. This set of NWs occupies a volume $V_2$. This approach was iterated, by applying a filter that removes areas with intensity similar to regions with two overlapping NWs (Figure S4d) and obtaining a volume $V_3$ of NWs. The NW/PMMA volumetric ratio can be calculated as the ratio between the sum $V_1+V_2+V_3$ and the volume of the polymer fiber, obtaining a value of about 6%. This is in line with the expected one, calculated starting from the experimental weight ratio used for sample preparation. In fact, considering the ZnO mass density, $D_{ZnO}$=5.4 g/cm$^3$ and the PMMA one, $D_{PMMA}$=1.19 g/cm$^3$, samples realized starting from a solution with $\chi = 30\%$, would feature a NW/PMMA volumetric ratio of 6.6%.

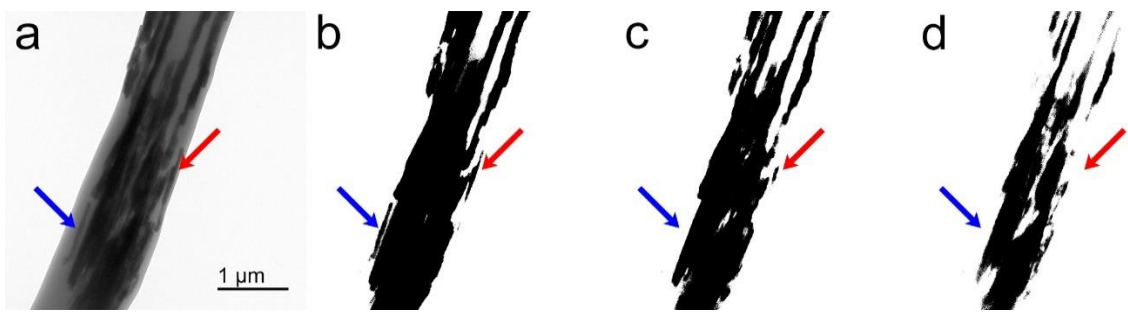

**Figure S4.** (a) TEM micrograph of a ZnO NW-in fiber hybrid ($\chi = 30\%$), used for the estimation of the volumetric ratio between NWs and PMMA matrix. (b-d) Micrographs converted into binary images by the application of a threshold which selects the regions (black areas) with grayscale intensity lower than 31% (b), 25.5% (c) and 19% (d) of the maximum intensity of the micrograph.

## 4. Optical properties

The effect of the fiber size and NW content on the light transport properties of the hybrids can be rationalized by considering a cylindrical waveguide with diameter, $D$, and refractive index, $n_{eff}$, and calculating the relative integrated intensities of the modes of electromagnetic field (with wavelength, $\lambda = 382$ nm, and wavenumber, $k = 2\pi/\lambda$) inside and outside the waveguide. This can be evaluated by calculating the fractional mode power, $\eta$, within the waveguide as follows:[1]

$$\eta = 1 - \frac{\left(2.405\, exp\left[-\frac{2}{kD\sqrt{n_{eff}-1}}\right]\right)^2}{\left(0.5kD\sqrt{n_{eff}-1}\right)^3} \tag{S1}$$

Figure S5 shows the dependence of $\eta$ on the waveguide diameter for a neat PMMA fiber ($n_{eff} = n_{PMMA} = 1.51$)[2] and for a fiber incorporating ZnO NWs ($\chi = 30\%$). For the hybrids, $n_{eff}$ was estimated by utilizing the Maxwell Garnett expression for the refractive index of composite materials, which contain anisotropic particles as fillers with length ($L_{NW}$) much larger than their transversal size ($L_{NW} >> d_{NW}$).[3] The volumetric ratio between the ZnO NW fillers and the PMMA matrix was 6%, as estimated above. The resulting value of $n_{eff}$ is 1.55. Figure S5 shows that for fibers with transversal size, $D$, larger than 1 μm, almost 99% of the field intensity is contained in the fiber, whereas a decrease of the fractional mode power occurs for $D < 250$ nm, namely close to the cut-off size, $d_{cut\text{-}off}$. The value of $d_{cut\text{-}off}$ is about 160 nm for the hybrid fibers (and about 170 nm for pristine PMMA fibers), as estimated by the expression:[1] $d_{cut\text{-}off} \cong \frac{\lambda}{2\sqrt{n_{eff}^2-1}}$.

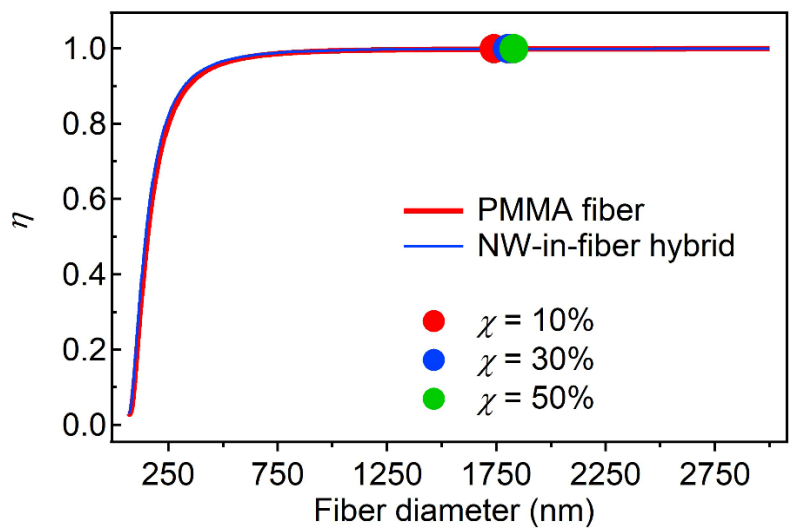

**Figure S5.** Calculated fractional mode power within a cylindrical waveguide, $\eta$, *vs.* the diameter of the fiber. The data are calculated for a neat PMMA fiber (red line, $n_{eff} = n_{PMMA} = 1.51$) and for a NW-in-fiber hybrid (blue line, $n_{eff} = 1.55$). The measured average diameter of the hybrid fibers obtained by electrospinning from solutions with $\chi = 10\%$, 30% and 50% ZnO NWs are marked by red, blue and green full circles, respectively.

The general behavior for light propagating in a complex waveguide constituted by a polymer matrix incorporating ZnO NWs can be captured taking into account the alignment of the NWs along the fiber longitudinal axis, as well as their elongated shape. Light can efficiently propagate in a fiber made of PMMA if the angle of incidence, $\beta$, at the fiber/air interface is larger than the critical angle ($\beta c = 41.5°$), as schematized in Figure S6a. Therefore, the light incident on a ZnO NW that is aligned along the fiber length will have local incidence angle, $\alpha_i$, in the range $\pm48.5°$.

Such beams will be then diffused by the NW. If the direction of the scattered light will not meet the condition for internal total reflection, this light will be scattered out of the fiber, whereas it will be waveguided in case of diffusion angles in a range ±48.5° with respect to the fiber axis (roughly parallel to NWs). To estimate the amount of light that fulfills this requirement, the angular dependence of the intensity of the light scattered by a NW is calculated under the Rayleigh-Gans approximation,[4] which is valid if $\left| \frac{n_{ZnO}}{n_{PMMA}} - 1 \right|$ is well below unity and $\frac{\pi d_{NW}}{\lambda} \left| \frac{n_{ZnO}}{n_{PMMA}} - 1 \right| \ll 1$, where $d_{NW}$ is the diameter of the ZnO NWs ($d_{NW}$=40 nm) and $n_{ZnO}$ is the refractive index of ZnO ($n_{ZnO} = 2.45$).[5] Under these conditions, the following expression can be used for the calculation of a form factor:[4]

$$f(\theta, \phi, \alpha_i)$$

$$= \frac{8 \sin\{0.5 k L_{NW}[\cos\alpha_i - \sin\theta\cos(\phi)]\}}{k L_{NW}[\cos\alpha_i - \sin\theta\cos(\phi)]} \frac{J_1(0.5 k d_{NW}\sqrt{\sin^2\theta\sin^2(\phi) + (\cos\theta - \sin\alpha_i)^2}}{k d_{NW}\sqrt{\sin^2\theta\sin^2(\phi) + (\cos\theta - \sin\alpha_i)^2}}$$

where $L_{NW}$ is the ZnO NW length, while the angles $\theta$, $\phi$ and $\alpha_i$ are introduced in Figure 6a. The intensity of the light scattered by the NW is proportional to the squared form factor[4] and, consequently, the calculation of $f^2(\theta, \phi)$ provides useful information about the angular distribution of intensity of light scattered by a NW embedded in PMMA. The results of such calculations for various angle of incidence are shown in Figure S6b-g, which evidences that most of the scattering occurs in the NW forward direction and in an angular range that allows for propagation along the fiber by total internal reflection.

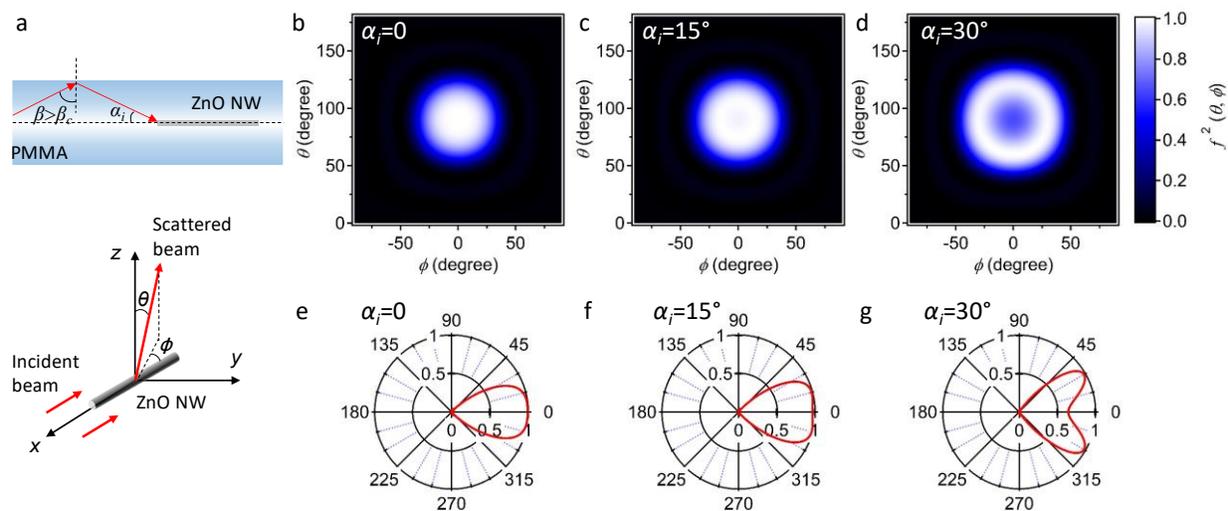

**Figure S6**. (a) Scheme of a light beam guided along a NW-in-fiber waveguide (top scheme) and of the light beam incident and scattered by an embedded ZnO NW (bottom scheme). (b-d) Dependence of the squared form factor, $f^2(\theta, \phi)$, on the scattering angles $\theta$ and $\phi$, that are schematized in the bottom scheme of (a), for different angles of incidence ($\alpha_i$) of the beam propagating in the fiber waveguide. (e-g) Corresponding dependence the squared form factor on $\phi$ in the $x$-$y$ plane ($\theta = 0$). In (e)-(g) the incident beam propagates from left to right, whereas $\phi = 0$ (180°) is the direction of forward(backward) scattering.

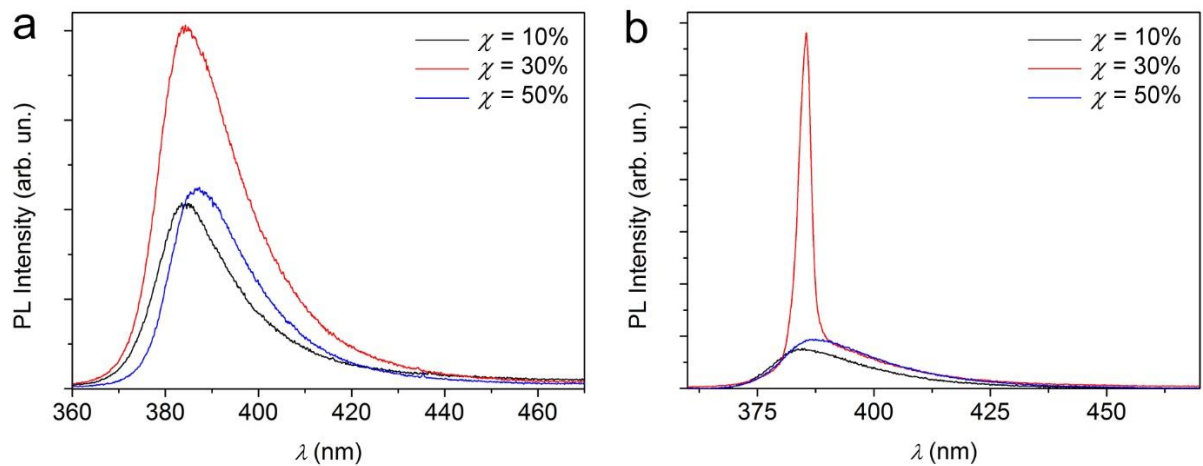

**Figure S7.** Comparison of PL emission (a) and ASE (b) in NW-in-fibers electrospun from solutions with $\chi$ = 10%, 30%, and 50%. Spectra are acquired with the same excitation and collection set-up. $\lambda_{exc}$ = 355 nm. Excitation fluence: 0.2 mJ cm$^{-2}$ (a) and 1.8 mJ cm$^{-2}$ (b).

The photoexcited carrier density has been estimated by considering that the pump pulse has an energy, $E_p$ = 20 nJ, and a beam diameter in the focus, $d$ = 100 μm, corresponding to a fluence $F$ = $E_p$ /(π$d^2$/4) ≅ 250 μJ/cm². Given the pump pulse photon energy of 3.7 eV (corresponding to 335 nm wavelength) we obtained an incident pump photon flux of 42.2×10¹³ photons/cm². Considering the optical density of the sample ($OD$ = 0.2 at 3.7 eV), and the corresponding absorbance A = 1-$10^{-OD}$ = 0.37, we obtained an absorbed photon flux of 15.6×10¹³ photons/cm². Finally, assuming a uniform carrier density across a sample thickness of 23 μm, which is a justified approximation given the comparatively low optical density, we finally estimated a photoexcited carrier density of 6.78×10¹⁶ carriers/cm³.

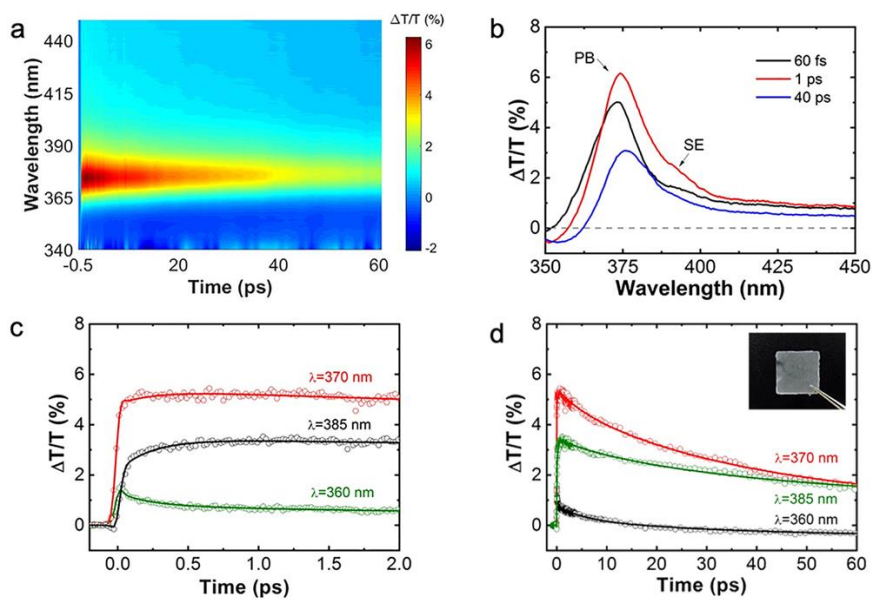

**Figure S8.** Femtosecond pump-probe spectroscopy of fibers with ZnO NPs. (a) 2D $\Delta T/T$ map as a function of probe wavelength and pump-probe delay. (b) $\Delta T/T$ spectra at selected pump-probe delays. (c,d) $\Delta T/T$ dynamics at selected probe wavelengths for short (c) and long (d) delays. Pump-probe measurements are performed on samples electrospun from solutions with $\chi$ = 10% -see picture in the inset of panel (d)-, analogously to Figure 5.

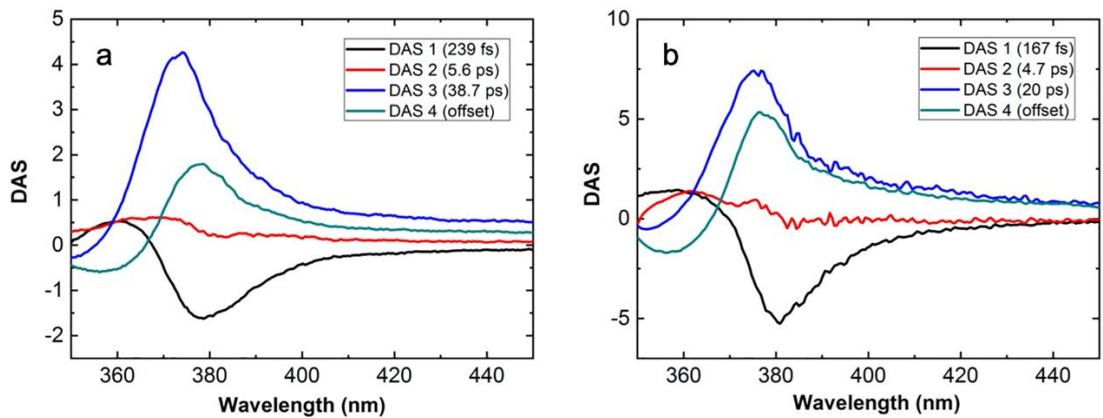

**Figure S9.** Decay associated spectra (DAS) for the femtosecond pump-probe measurements on the fibers with ZnO NWs (a) and NPs (b).

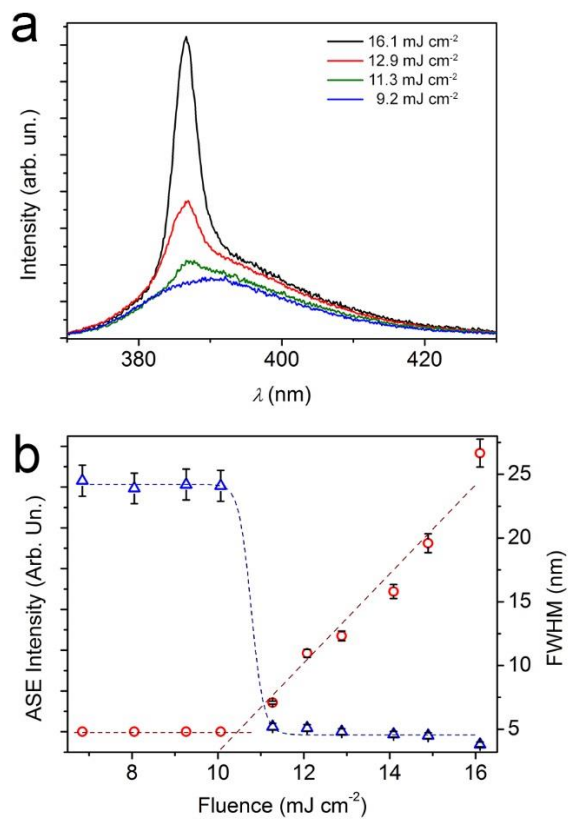

**Figure S10.** (a) ASE spectra of NW-doped films for different pumping fluences. (b) ASE intensity (red circles, left vertical scale) and full width at half maximum (FWHM) of the spectra (blue triangles, right vertical scale) as a function of the pumping fluence.

| UV emitter | ASE/ Laser | Max. emission wavelength (nm) | Threshold (mJ cm$^{-2}$) | Threshold (kW cm$^{-2}$) | FWHM (nm) | Polarization | Divergence (mrad) | Flexible/ Conformable | Ref. |
|---|---|---|---|---|---|---|---|---|---|
| ZnO NWs-in-fiber hybrids | ASE | 382 | 0.6 | 60 | 4 | Polarized | 5 | Yes | This work |
| Polymer fibers doped with organic dye | ASE | 387 | 0.1 | 10 | 11 | - | 16.5 | Yes | 6 |
| ZnO NPs - polymer film | Laser | 387 | 2.8 | - | 4 | - | - | Yes | 7 |
| Organic semicon-ductor film | ASE | 394 | $1.3 \times 10^{-3}$ | 2.6 | 3.6 | - | - | - | 8 |
| ZnO NWs array | Laser | 385 | - | 40 | <0.3 | - | - | - | 9 |
| ZnO NW | Laser | 385 | $7 \times 10^{-5}$ | - | 0.25-1 | Polarized | - | - | 5 |
| Hexagonal ZnO Microdisks | Laser | 390 | 0.28 | 280 | 0.12 | - | - | - | 10 |
| GaN NW | Laser | 375 | $7 \times 10^{-4}$ | - | <1 | - | - | - | 11 |
| GaN NW /silica layer/Al film | Laser | 370 | - | 3500 | 0.8 | Polarized | - | - | 12 |
| 2,4,5-triphenyl-imidazole NW | Laser | 375 | - | - | <4 | - | - | - | 13 |

**Table S1.** Comparison of the emission properties of ZnO NW-in-fiber hybrids (this work) and other UV-emitting materials which show amplified emission and lasing under optical pumping.